\documentclass[seceq]{ptptex}
\usepackage{graphicx} 
\usepackage{latexsym}
\notypesetlogo
\begin{document}
  \begin{flushright}
   RUP-07-2 \\
   October, 2007
\end{flushright}
\vspace{10mm} 

\begin{center}
\Large{\bf  Kinematical Constraints on QCD Factorization \\ in the Drell-Yan  Process}

\vspace{15mm}

\large{Hidekazu {\sc Tanaka}, Yu {\sc Matsuda} and Hirokazu {\sc Kobayashi} } \\
Department of Physics, Rikkyo University, \\
           Nishi-ikebukuro, Toshima-ku Tokyo, Japan, 171 \\


\vspace{25mm}

  {\Large ABSTRACT}
  \end{center}
        
\vspace{10mm}

We study factorization schemes for parton shower models in hadron-hadron collisions.  As an example, we calculate lepton pair production mediated by a virtual photon in quark--anti-quark annihilation, and we compare factorized cross sections obtained in the conventional $\overline{\rm MS}$ scheme with those obtained in a factorization scheme in which a kinematical constraint due to parton radiation is taken into account.   We discuss some properties of factorized cross sections.


\section{Introduction}

Hadron-hadron scattering processes are useful for investigating various properties of high-energy reactions, as well as to search for exotic particles. 
 However, the leading-logarithmic (LL) order of quantum chromodynamics (QCD) is insufficient to evaluate these processes.  Thus the next-to-leading logarithmic (NLL) order contributions should be taken into account.

In the evaluation of hadron scattering cross sections, logarithmic contributions due to collinear parton production are subtracted from a hard scattering cross section, and they are absorbed into the parton distributions of hadrons.  The next-to-leading-order (NLO) calculation is necessary in order to remove theoretical ambiguities due to the factorization procedure as well as the choice of the factorization scale. 

In actual calculations, Monte Carlo methods are powerful tools for the evaluation of exclusive processes. 
The initial state parton radiation can be generated by using the parton shower models based on perturbative QCD. One such algorithm is proposed in Ref.\citen{rf:1}, in which the parton showers are generated according to an algorithm consisting of a model based on the evolution of momentum distributions with the modified minimal subtraction ($\overline{\rm MS}$) scheme.\cite{rf:2}   In this model, the scaling violation of the parton distributions is generated with only information from the splitting functions of the parton branching vertices and input distributions at a given low energy.   
  It has been found that this method reproduces the scaling violation of the flavor singlet parton distributions, up to their normalizations, to an accuracy of the NLL order of QCD.  
 Furthermore, exact NLL-order vertex functions for the decay branching processes are taken into account in this model.\cite{rf:3}

  In this paper, we study factorization schemes for the parton shower models in hadron-hadron collisions, in which the collinear singularities are subtracted using the $\overline{\rm MS}$ scheme.  In order to implement factorization schemes that are appropriate for the Monte Carlo methods using the parton shower models to an accuracy of the NLL order of QCD, we take into account a kinematical constraint due to virtual partons.  The momenta of partons in the scattering processes are conserved in this method.

At NLO accuracy, the processes $q{\bar q}\rightarrow \gamma^*g$, $qg\rightarrow \gamma^*q$ and ${\bar q}g\rightarrow \gamma^*{\bar q}$ contribute to the hard scattering cross section in actual physical systems.  As an example, we calculate  factorized cross sections for the process  $q{\bar q}\rightarrow \gamma^*g$  using our method. Other processes can be calculated with similar method presented in this paper. 
 
 In $\S$2, we summarize the calculation of the cross section of the $q{\bar q}\rightarrow \gamma^*g$ process using the dimensional regularization method.  Factorization schemes for the collinear singularity are explained in $\S$3. Some properties of the factorized cross sections are presented in $\S$4.
Section 5 contains a summary and some comments.

\section{Cross section of  the Drell-Yan process}

In this section, we summarize the calculation of the cross section of the Drell-Yan lepton-pair production in quark $(q)$--anti-quark $({\bar q})$ annihilation,
\begin{eqnarray}
 q(p_q)+{\bar q}(p_{\bar q}) \rightarrow \gamma^*(q)+g(p_g) \rightarrow \l^-(p_-)+l^+(p_+)+g(p_g), 
 \end{eqnarray}
mediated by a photon $\gamma^*$ with virtuality $(p_- + p_+)^2=q^2=Q^2$, where a gluon $(g)$ is radiated in the final state. Here, $p_i~(i=q,{\bar q},g)$ and $p_{\pm}$ denote the momenta of the corresponding particles.

Ignoring the particle masses, the cross section of the Born term,  
\begin{eqnarray}
 q(p_q)+{\bar q}(p_{\bar q}) \rightarrow \gamma^*(q) \rightarrow l^-(p_-)+l^+(p_+), 
 \end{eqnarray}
is given by
\begin{eqnarray}
{d{\hat \sigma}^{(B)}_{q{\bar q}} \over dQ^2}={\hat \sigma}_0(Q^2,\epsilon)\delta({\hat s}-Q^2),
 \end{eqnarray}
 with
\begin{eqnarray}
 {\hat \sigma}_0(Q^2,\epsilon)={\pi\alpha \over N_CQ^2}{\hat e}_q^2(2-2\epsilon){2\alpha \over 3-2\epsilon}{\Gamma(2-\epsilon) \over \Gamma(2-2\epsilon)}\left[{Q^2 \over 4\pi\mu^2}\right]^{-\epsilon},
  \end{eqnarray}
 in $4-2\epsilon$ dimensions. Here $N_C=3$ and the electric coupling constant of the quark is defined by ${\hat e}_q^2\alpha\mu^{2\epsilon}$, with the mass parameter $\mu$ and the dimensionless coupling $\alpha$.  

The differential cross section for real gluon radiation is given by
 \begin{eqnarray}
{d{\hat \sigma}^{(r)}_{q{\bar q}} \over d{\hat \tau}d(-{\hat t})}={\alpha_s \over 2\pi{\hat s}} {\hat \sigma}_0(Q^2,\epsilon)\left[{(-{\hat t})(-{\hat u}) \over 4\pi{\hat s}\mu^2}\right]^{-\epsilon}{1 \over \Gamma(1-\epsilon)}{\tilde K}_{q{\bar q}}^{(r)}({\hat \tau},-{\hat t},\epsilon), 
\end{eqnarray}
 with
 \begin{eqnarray}
 {\tilde K}_{q{\bar q}}^{(r)}({\hat \tau},-{\hat t},\epsilon)=C_F\left[{\hat s}\left({1 \over -{\hat t}}+{1 \over -{\hat u}}\right){1 \over C_F}{\hat P}_{qq}({\hat \tau},\epsilon)-2\right], 
\end{eqnarray}
where ${\hat \tau}=Q^2/{\hat s}$ and  
\begin{eqnarray}
  {\hat P}_{qq}({\hat \tau},\epsilon)={\hat P}_{qq}^{(0)}({\hat \tau})+\epsilon {\hat P}'_{qq}({\hat \tau}),
 \end{eqnarray}
 with
\begin{eqnarray}
  {\hat P}_{qq}^{(0)}({\hat \tau})=C_F{1+{\hat \tau}^2 \over 1-{\hat \tau}} , ~~  {\hat P}'_{qq}({\hat \tau})=-C_F(1-{\hat \tau}). 
\end{eqnarray}
 Here, $C_F=4/3$ and $\alpha_s$ are the color factor and the strong coupling constant, respectively.   The Mandelstam variables are defined by
\begin{eqnarray}
{\hat s}=(p_q+p_{\bar q})^2,~~{\hat t}=(p_q-p_g)^2, ~~{\hat u}=(p_{\bar q}-p_g)^2,
\end{eqnarray}
which satisfy ${\hat s}+{\hat t}+{\hat u}=Q^2$.

Integrating over  the range $0 \leq -{\hat t} \leq {\hat s}(1-{\hat \tau})$ and adding virtual loop contributions,  we obtain the cross section \cite{rf:4} as
\begin{eqnarray}
{d{\hat \sigma}_{q{\bar q}} \over d{\hat \tau}}={\alpha_s \over 2\pi}{\hat \sigma}_0(Q^2,\epsilon){\Gamma(1-\epsilon) \over \Gamma(1-2\epsilon)}\left[{{\hat s} \over 4\pi\mu^2}\right]^{-\epsilon}K_{q{\bar q}}({\hat \tau},\epsilon),
\end{eqnarray}
with
\begin{eqnarray}
K_{q{\bar q}}({\hat \tau},\epsilon)&=&C_F\Big[{2 \over -\epsilon}{1 \over C_F}\left({\hat P}_{qq}^{(0)}({\hat \tau})\right)_++4(1+{\hat \tau}^2)\left({\log(1-{\hat \tau}) \over 1-{\hat \tau}}\right)_+ \nonumber \\
&+&\left(-8+{2 \over 3}\pi^2\right)\delta(1-{\hat \tau})\Big].
\end{eqnarray}

Here, we have 
\begin{eqnarray}
 \left({\hat f}({\hat \tau})\right)_+={\hat f}({\hat \tau})-\delta(1-{\hat \tau})\int^1_0dy{\hat f}(y) 
\end{eqnarray}
for a function ${\hat f}({\hat \tau})$ unregulated at ${\hat \tau}=1$.

\section{Factorization schemes}

In actual Monte-Carlo simulations with the parton shower models, the virtualities and longitudinal momentum fractions of the partons in the initial state are generated according to non-branching probabilities (form factors) and splitting functions of parton branching processes, respectively.\cite{rf:1}  The four-momenta of the partons are constructed from these values.

In order to obtain a finite cross section for the process $q{\bar q}\rightarrow \gamma^*g$, we subtract the collinear contributions corresponding to the branching processes $q \rightarrow qg$ and ${\bar q} \rightarrow {\bar q}g$ from the hard scattering cross section.

The momentum of the quark $r_q$, defined by $r_q=p_q-p_g$ for the branching process
\begin{eqnarray}
q(p_q) \rightarrow q(r_q)+g(p_g),
\end{eqnarray}
is described by using the momentum fraction $z_q$, the virtuality $r_q^2$, and the transverse momentum $r_{qT}$, with $p_q\cdot r_{qT}=p_{\bar q}\cdot r_{qT}=0$, as 
\begin{eqnarray}
 r_q=p_q-p_g=z_qp_q+\alpha_q p_{\bar q}+r_{qT}, 
\end{eqnarray}
where $\alpha_q=r_q^2/{\hat s}$ for on-shell gluon radiation ($p_g^2=0$). 
Here, we set $p_q^2=p_{\bar q}^2=0$, because the relations $-p_q^2,-p_{\bar q}^2 \ll -r^2_q$ are expected in parton shower generation.

The virtuality of a photon $Q^2$ is expressed as  
\begin{eqnarray}
Q^2=(r_q+p_{\bar q})^2,
\end{eqnarray}
to $O(\alpha_s)$ accuracy of QCD.  Using Eqs. (3$\cdot$2) and (3$\cdot$3), the quantity ${\hat \tau}$ is given by
\begin{eqnarray}
{\hat \tau}=z_q+\alpha_q.
\end{eqnarray}

According to Eq. (3$\cdot$4), the subtraction term  divided by ${\hat \sigma}_0(Q^2,\epsilon)$  is defined by
\begin{eqnarray}
 {dS_{qq}^{[F]} \over d{\hat \tau}d(-r^2_q)}=\int^1_0dz_q{d{\tilde S}_{qq}^{[F]} \over dz_q d(-r^2_q)}\delta(z_q-{\hat \tau}-(-r^2_q)/{\hat s}),
 \end{eqnarray}
with
 \begin{eqnarray}
 {d{\tilde S}_{qq}^{[F]} \over dz_q d(-r^2_q)}={\alpha_s \over 2\pi}{1 \over \Gamma(1-\epsilon)}\left[{-r^2_q \over 4\pi\mu^2}\right]^{-\epsilon}{P_{qq}^{[F]}(z_q,\epsilon) \over -r^2_q},
 \end{eqnarray}
 where the splitting function $P_{qq}^{[F]}(z_q,\epsilon)$ depends on the factorization scheme $F$. 

The conventional $\overline{\rm MS}$ subtraction scheme\cite{rf:5} corresponds to the approximation $ \delta(z_q-{\hat \tau}-(-r^2_q)/{\hat s}) \simeq \delta(z_q-{\hat \tau})$ in Eq. (3$\cdot$5). In this approximation, the collinear contribution is subtracted from the hard scattering cross section in the range $ 0 \leq -{\hat t} \leq M^2$, with ${\hat \tau}=z_q$ and $-{\hat t}=-r^2_q$. Here, the mass parameter $M$ corresponds to a factorization scale for the separation between the initial state radiation and the hard scattering process. In the case of the conventional $\overline{\rm MS}$ scheme, $M^2$ is larger than the kinematical boundary for $-{\hat t}$, given by ${\hat s}(1-{\hat \tau})$ in $1-{\hat \tau}_M \leq {\hat \tau} \leq 1$, where we define ${\hat \tau}_M \equiv M^2/{\hat s}$.  Taking the condition $z_q={\hat \tau}+(-r^2_q)/{\hat s} \leq 1$ into account, the collinear contributions are subtracted within the phase space for the hard scattering process, namely $0 \leq -{\hat t} \leq {\hat s}(1-{\hat \tau}) $.

We define the integrated contribution as 
\begin{eqnarray}
 \int_0^{{\hat s}w^{[F](I)}} d(-r^2_q){dS_{qq}^{[F]} \over d{\hat \tau}d(-r^2_q)}= {\alpha_s \over 2\pi}{1 \over \Gamma(1-\epsilon)}\left[{{\hat s} \over 4\pi\mu^2}\right]^{-\epsilon}{\tilde F}_{qq}^{[F](I)}(\epsilon,{\hat \tau}),
 \end{eqnarray}
where $I$ denotes the region of phase space for the hard scattering process to be considered. Here, ${\hat s}w^{[F](I)}$ is a boundary of the $-r^2_q$ integration.
For example, the splitting function  with the $\overline{\rm MS}$ scheme is defined by 
 \begin{eqnarray}
  P_{qq}^{[\overline{\rm MS}]}(z_q,\epsilon)&=&\left({\hat P}_{qq}^{(0)}(z_q)\right)_+,
   \end{eqnarray}
  for the branching process presented in Eq. (3$\cdot$1).
 
With $z_q={\hat \tau}(F=\overline{\rm MS})$ and $w^{[F](S)}={\hat \tau}_M$, the subtraction term  for  collinear gluon radiation is given by  
\begin{eqnarray}
 {\tilde F}^{[\overline{\rm MS}](S)}_{qq}(\epsilon,{\hat \tau}) =  \left({1\over -\epsilon}+\log{\hat \tau}_M\right) P_{qq}^{[\overline{\rm MS}]}({\hat \tau},\epsilon).
\end{eqnarray}
Similarly, with $z_q={\hat \tau}+(-r^2_q)/{\hat s}$ ($F=\overline{\rm MS}'$) and $w^{[\overline{\rm MS}'](S)}=1-{\hat \tau}$, we obtain 
\begin{eqnarray}
 {\tilde F}^{[\overline{\rm MS}'](S)}_{qq}(\epsilon,{\hat \tau}) &=& C_F\Big[{1\over -\epsilon}{1 \over C_F}\left({\hat P}_{qq}^{(0)}({\hat \tau})\right)_++(3+{\hat \tau}^2)\left({\log(1-{\hat \tau}) \over 1-{\hat \tau}}\right)_+  \nonumber \\
& & +{3 \over 2}{1 \over (1-{\hat \tau})_+}-(1-{\hat \tau})-{1 \over 3}\pi^2\delta(1-{\hat \tau})\Big],
\end{eqnarray}
which is valid for $1-{\hat \tau}_M \leq {\hat \tau} \leq 1$ ($I=S$). 

The calculational method employed in the soft gluon region is explained in Appendix A.  By replacing $q$ with ${\bar q}$ in the above equations, we obtain the subtraction terms for the anti-quark legs, which are the same as those for the quark legs.

\section{Factorized cross sections}

The factorized cross section in the factorization scheme $F$ is defined by
\begin{eqnarray}
{d{\hat \sigma}_{q{\bar q}}^{[F](S)} \over d{\hat \tau}}=
{\alpha_s \over 2\pi}{\hat \sigma}_0(Q^2,0)K_{q{\bar q}}^{[F](S)}({\hat \tau}),
\end{eqnarray}
with 
\begin{eqnarray}
K_{q{\bar q}}^{[F](S)}({\hat \tau})=K_{q{\bar q}}({\hat \tau},\epsilon)-{\tilde F}^{[F](S)}_{qq}(\epsilon,{\hat \tau})-{\tilde F}^{[F](S)}_{{\bar q}{\bar q}}(\epsilon,{\hat \tau}),
\end{eqnarray}
for $1-{\hat \tau}_M \leq {\hat \tau} \leq 1$. Here, we obtain
\begin{eqnarray}
K_{q{\bar q}}^{[\overline{\rm MS}](S)}({\hat \tau})&=&C_F\Big[4(1+{\hat \tau}^2)\left({\log(1-{\hat \tau}) \over 1-{\hat \tau}}\right)_+ \nonumber \\
&-& {2 \over C_F}\left({\hat P}_{qq}^{(0)}({\hat \tau})\right)_+\log{\hat \tau}_M + \left(-8+{2 \over 3}\pi^2\right)\delta(1-{\hat \tau})\Big]
\end{eqnarray}
with the $\overline{\rm MS}$ scheme,\cite{rf:5} and 
\begin{eqnarray}
K_{q{\bar q}}^{[\overline{\rm MS}'](S)}({\hat \tau})&=&C_F\Big[-2(1+{\hat \tau})\log(1-{\hat \tau}) -{3 \over (1-{\hat \tau})_+}+2(1-{\hat \tau}) \nonumber \\
&+& \left(-8+{4 \over 3}\pi^2\right)\delta(1-{\hat \tau})\Big]
\end{eqnarray}
with the $\overline{\rm MS}'$ scheme.

The factorized cross section in the $\overline{\rm MS}$ scheme depends on the factorization scale $M$. With the $\overline{\rm MS}'$ scheme, the factorized cross section has no $M$ dependence, because the subtraction term is integrated over the range $0 \leq -r_q^2 \leq {\hat s}(1-{\hat \tau})$. 

In order to evaluate the soft gluon contribution, we integrate Eq. (4$\cdot$1) over the range $(1-\eta_s){\hat s} \leq Q^2 \leq {\hat s}$, with fixed ${\hat s}$, where ${\hat s}$ is generated by the parton showers. Here, $\eta_s$ is a cut-off parameter, satisfying $\eta_s \leq {\hat \tau}_M$. 
 The energy scale of the running coupling constant in this region can be chosen as $Q^2$. In order to simplify our analysis, we evaluate the soft gluon contribution with the coupling constant $\alpha_s(Q^2)\simeq\alpha_s({\hat s})$.  The running coupling constant to the accuracy of the NLL order is normalized as $\alpha_s(M_Z^2)=0.114$ at the $Z^0$ boson mass.\cite{rf:6}

The integrated cross section is given by 
\begin{eqnarray}
\sigma_{q{\bar q}}^{[F](S)} (1,1-\eta_s)\simeq{\alpha_s({\hat s}) \over 2\pi}{\hat \sigma}_0({\hat s},0)I_{q{\bar q}}^{[F](S)} (1,1-\eta_s),
\end{eqnarray}
with
\begin{eqnarray}
I_{q{\bar q}}^{[F](S)} (1,1-\eta_s)=\int^1_{1-\eta_s} {d{\hat \tau} \over {\hat \tau}}K^{[F](S)}_{q{\bar q}}({\hat \tau}).
\end{eqnarray}

Integrating over the range $1-\eta_s \leq {\hat \tau} \leq 1$, we obtain 
\begin{eqnarray}
I_{q{\bar q}}^{[\overline{\rm MS}](S)}(1,1&-&\eta_s)= C_F\Big[4\left\{SP_-(\eta_s,0)-\eta_s(\log\eta_s-1)+\log^2\eta_s\right\} \nonumber \\
& & -2\left\{-\log(1-\eta_s)-\eta_s+{3 \over 2}+2\log\eta_s\right\}\log{\hat \tau}_M-8+{2 \over 3}\pi^2\Big] 
\end{eqnarray}
with the $\overline{\rm MS}$ scheme, and 
\begin{eqnarray}
I_{q{\bar q}}^{[\overline{\rm MS}'](S)}(1,1-\eta_s)&=& C_F\Big[-2SP_-(\eta_s,0)-(2\eta_s+3)\log\eta_s \nonumber \\
& & +\log(1-\eta_s)-8+{4 \over 3}\pi^2\Big] 
\end{eqnarray}
with the $\overline{\rm MS}'$ scheme.  Here, the function $SP_-$ is defined by 
\begin{eqnarray}
SP_{-}(a,b) \equiv \int^a_bd{\hat \tau}{\log{\hat \tau} \over 1 - {\hat \tau}}.
\end{eqnarray}

We calculate the $\eta_s$ dependence of the function 
\begin{eqnarray}
R^{[F](S)}(\eta_s)=1+{\sigma_{q{\bar q}}^{[F](S)} (1,1-\eta_s) \over {\hat \sigma}_0({\hat s},0)}\simeq1+{\alpha_s({\hat s}) \over 2\pi}I_{q{\bar q}}^{[F](S)} (1,1-\eta_s)
\end{eqnarray}
by using the expressions appearing in Eqs. (4$\cdot$7) and (4$\cdot$8).

In Fig.1, the $\eta_s$ dependence of the contribution obtained with the $\overline{\rm MS}'$ scheme is plotted by the solid curve.   The result obtained with the $\overline{\rm MS}$ scheme for $\eta_s={\hat \tau}_M \leq 0.5$, which corresponds to the contribution integrated over the range $1-{\hat \tau}_M \leq {\hat \tau} \leq 1$, is represented by the dash-dotted curve.   The $\eta_s$ dependences found with the two schemes become similar, because the $\log^2\eta_s$ term is canceled by the term $\log\eta_s\log{\hat \tau}_M$ in the $\overline{\rm MS}$ scheme, represented by Eq. (4$\cdot$7).  

 The dashed curve plots the $\eta_s$ dependence obtained with the $\overline{\rm MS}$ scheme at ${\hat \tau}_M=0.5$.  The cross section near the threshold ($\eta_s \ll {\hat \tau}_M$) becomes large due to the $\log^2\eta_s$ term in Eq. (4$\cdot$7).  The error in the calculation is approximated by $1 \leq \alpha_s((1-\eta_s){\hat s})/\alpha_s({\hat s}) < 1.05$ for $0 \leq \eta_s \leq {\hat \tau}_M=0.5$ at $\sqrt{\hat s}=200~{\rm GeV}$.
\begin{figure}
\centerline{\includegraphics[width=10cm]{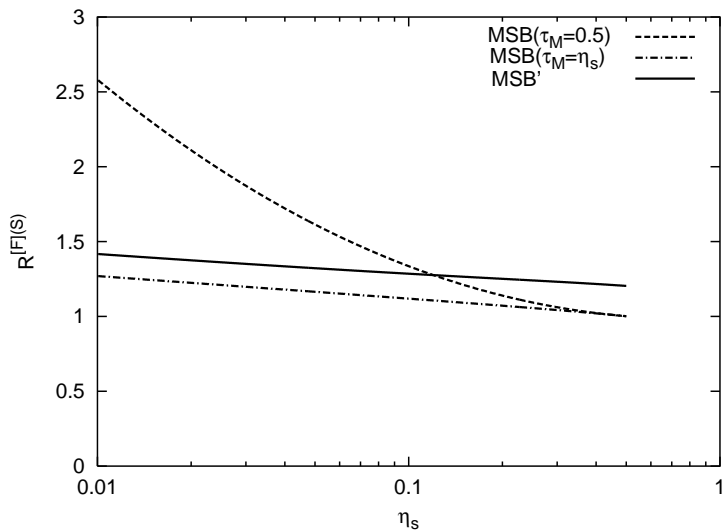}}
\caption{The integrated contributions of the cross sections divided by ${\hat \sigma}_0({\hat s},0)$, which are defined by Eq. (4$\cdot$10), with $\sqrt{\hat s}=200~{\rm GeV}$.  The solid curve represents the result obtained with the $\overline{\rm MS}'$ scheme.  With the $\overline{\rm MS}$ scheme, the dashed curve and the dash-dotted curve represent the result found with ${\hat \tau}_M=0.5$ and that found with ${\hat \tau}_M=\eta_s$, respectively.  
}
\end{figure}

 Next, we calculate the factorized cross sections for hard and collinear gluon radiation ($I=C$) with ${\hat \tau}_M \ll 1-{\hat \tau}$ (see Appendix B).
The remnant in the hard collinear region integrated over the range $0 \leq -{\hat t} \leq M^2$ is given by
\begin{eqnarray}
 K_{q{\bar q}}^{[\overline{\rm MS}](C)}({\hat \tau})\simeq K_{q{\bar q}}^{[\overline{\rm MS}'](C)}({\hat \tau}) \simeq C_F\left[{1 \over C_F}{\hat P}^{(0)}_{qq}({\hat \tau})\log(1-{\hat \tau})+(1-{\hat \tau})\right]. 
\end{eqnarray}
Negative contributions remain for ${\hat \tau} \sim 1$ both for the cross section obtained with the $\overline{\rm MS}$ scheme and that obtained with the $\overline{\rm MS}'$ scheme.

Though the parton showers already include the collinear contributions of the $g \rightarrow qX$ and $g \rightarrow {\bar q}X$ processes, as well as  those of the $q \rightarrow qX$ and ${\bar q} \rightarrow {\bar q}X$ processes, at NLL order\cite{rf:1}, we have to add the NLO contributions for the processes $qg \rightarrow \gamma^*q$ and ${\bar q}g \rightarrow \gamma^*{\bar q}$ in the hard scattering region. 
The calculation of the factorized cross sections is rather straightforward, because there is no infra-red singularity at ${\hat \tau} = 1$.

\section{Summary and comments}

In this paper, we have studied factorization schemes for hard scattering cross sections in hadron-hadron collisions, in which initial state radiation is generated by parton shower models.  Here, a kinematical constraint due to virtual parton contributions is taken into account in the subtraction of collinear singularities. In this method, this kinematical constraint guarantees the proper phase space boundary for the subtraction terms.  

As an example, we calculated hard scattering cross sections for the process $q{\bar q} \rightarrow \gamma^*g$ using the kinematical constraint for the gluon radiation. This method is called the $\overline{\rm MS}'$ scheme. We also calculated hard scattering cross section for this process without using the kinematical constraint.

The conventional $\overline{\rm MS}$ scheme gives a double logarithmic term, which increases the cross section in the soft gluon limit, whereas the $\overline{\rm MS}'$ scheme gives only a single logarithmic term. The infra-red behavior is rather stable in the case of the factorization with the $\overline{\rm MS}'$ scheme.  

In the collinear region $0 \leq -{\hat t} \leq M^2$,  after the collinear singularity is subtracted, negative contributions remain for ${\hat \tau} \sim 1$ both for the cross section obtained with the $\overline{\rm MS}$ scheme and that obtained with the $\overline{\rm MS}'$ scheme.  Such contributions cannot be ignored at NLL-order accuracy. However, event generations employed in the Monte Carlo methods according to the negative probability may not be appropriate.  We will discuss this point in future papers.  

Realizing the matching between the initial parton radiation generated by using the parton showers at NLL order and the hard scattering cross sections at NLO is not a trivial problem.  Further studies of this problem are necessary in order to construct consistent algorithms to evaluate higher-order contributions in QCD.

\section*{Acknowledgements}

This work was supported in part by the Rikkyo University Special Fund for Research and the Japan Society for the Promotion of Science through a Grant-in-Aid for Scientific Research(B)(No. 17340085).

\vspace{5mm}

\begin{center}
{\Large Appendix A}
\end{center}
\vspace{5mm}

In order to evaluate the quantity
\begin{eqnarray}
{\tilde F}_{qq}^{[\overline{\rm MS}'](S)}(\epsilon,{\hat \tau})={\hat s}^{\epsilon}\int^{{\hat s}(1-{\hat \tau})}_0d(-r^2_q)(-r^2_q)^{-1-\epsilon}\int^1_0dz_q\delta(z_q-{\hat \tau}-(-r^2_q)/{\hat s})C_F\left({1+z_q^2 \over 1-z_q}\right)_+, \nonumber
\end{eqnarray}
we calculate
\begin{eqnarray*}
{\tilde G}_{qq}^{[\overline{\rm MS}'](S)}(\epsilon,{\hat \tau})&=&\int^1_{\hat \tau}dy{\tilde F}_{qq}^{[\overline{\rm MS}'](S)}(\epsilon,y)=C_F\Big[{1 \over -\epsilon}\left\{2\log(1-{\hat \tau})+ {\hat \tau}+{1 \over 2}{\hat \tau}^2\right\}+2\log^2(1-{\hat \tau}) \nonumber \\
& & +\left({\hat \tau}+{1 \over 2}{\hat \tau}^2\right)\log(1-{\hat \tau})+1-{\hat \tau}^2+{1 \over 4}(1-{\hat \tau})^2-{\pi^2 \over 3}\Big]. 
\end{eqnarray*}
The function ${\tilde F}_{qq}^{[\overline{\rm MS}'](S)}(\epsilon,{\hat \tau})$ is given by
\begin{eqnarray*}
{\tilde F}_{qq}^{[\overline{\rm MS}'](S)}(\epsilon,{\hat \tau})&=&-\left({d{\tilde G}_{qq}^{[\overline{\rm MS}'](S)}(\epsilon,{\hat \tau}) \over d{\hat \tau}}\right)_++\delta(1-{\hat \tau}){\tilde G}_{qq}^{[\overline{\rm MS}'](S)}(0) \nonumber \\
&=&C_F\Big[{1 \over -\epsilon}{1 \over C_F}\left({\hat P}_{qq}^{(0)}({\hat \tau})\right)_++(3+{\hat \tau}^2)\left({\log(1-{\hat \tau}) \over 1-{\hat \tau}}\right)_+  \nonumber \\
& & -(1-{\hat \tau})+{3 \over 2}{1 \over (1-{\hat \tau})_+}-{1 \over 3}\pi^2\delta(1-{\hat \tau})\Big],
\end{eqnarray*}
which is presented in Eq. (3$\cdot$10) of the main text. 

\vspace{5mm}

\begin{center}
{\Large Appendix B}
\end{center}
\vspace{5mm}

In this appendix, we derive a remnant in the hard collinear region ($I=C$) for ${\hat \tau}_M \ll 1-{\hat \tau}$. 
The function $ {\tilde K}_{q{\bar q}}^{(r)}({\hat \tau},-{\hat t},\epsilon)$ presented in Eq. (2$\cdot$6) integrated over the range $0 \leq -{\hat t} \leq M^2$ ( or  $0 \leq -{\hat u} \leq M^2$) is given by 
\begin{eqnarray*}
& & K_{q{\bar q}}^{(C)}({\hat \tau},\epsilon)={1 \over {\hat s}}\int^{M^2}_0d(-{\hat t}){\tilde K}_{q{\bar q}}^{(r)}({\hat \tau},-{\hat t},\epsilon)  \nonumber \\
& & ~~~~~\simeq  C_F\Big[{1 \over C_F}{\hat P}_{qq}^{(0)}({\hat \tau})\left({1 \over -\epsilon}+   \log{\hat \tau}_M(1-{\hat \tau})\right)+1-{\hat \tau}\Big].
\end{eqnarray*}
The factorization terms are given by  
\begin{eqnarray*}
 {\tilde F}^{[\overline{\rm MS}](C)}_{qq}(\epsilon,{\hat \tau}) = \left({1\over -\epsilon}+\log{\hat \tau}_M\right) {\hat P}^{(0)}_{qq}({\hat \tau}) \simeq {\tilde F}^{[\overline{\rm MS}'](C)}_{qq}(\epsilon,{\hat \tau}).
\end{eqnarray*}
Thus we obtain the remnant in the hard collinear region as
\begin{eqnarray*}
 K_{q{\bar q}}^{[\overline{\rm MS}](C)}({\hat \tau})&=&K_{q{\bar q}}^{(C)}({\hat \tau},\epsilon)-{\tilde F}^{[\overline{\rm MS}](C)}_{qq}(\epsilon,{\hat \tau})\simeq K_{q{\bar q}}^{[\overline{\rm MS}'](C)}({\hat \tau}) \\ 
&\simeq& C_F\left[{1 \over C_F}{\hat P}^{(0)}_{qq}({\hat \tau})\log(1-{\hat \tau})+(1-{\hat \tau})\right],
\end{eqnarray*}
which is presented in Eq. (4$\cdot$11) of  the main text. 

The contribution of the hard gluon radiation ($I=H$) integrated over the range $M^2 \leq -{\hat t} \leq {\hat s}(1-{\hat \tau})-M^2$ is given by 
\begin{eqnarray*}
 K^{(H)}_{q{\bar q}}({\hat \tau})&=&{1 \over {\hat s}}\int^{{\hat s}(1-{\hat \tau})-M^2}_{M^2}d(-{\hat t}){\tilde K}_{q{\bar q}}^{(r)}({\hat \tau},-{\hat t},\epsilon)  \nonumber \\
&=& 2C_F\left[{1 \over C_F} {\hat P}_{qq}^{(0)}({\hat \tau})\log{1-{\hat \tau}-{\hat \tau}_M \over {\hat \tau}_M}-(1-{\hat \tau}-2{\hat \tau}_M)\right], 
\end{eqnarray*}
for $ {\hat \tau} \leq 1-2{\hat \tau}_M$.

\end{document}